\documentclass[useAMS,usenatbib]{mn2e}
\usepackage{graphicx}
\usepackage{color}
\usepackage{ulem}
\def\apj{ApJ}
\def\apjs{ApJS}

\def\aap{A\&A}

\def\mnras{MNRAS}

\def\pasa{PASA}

\def\nat{Nature}

\def\rmxaa{Revista Mexicana de Astronomia y Astrofisica}

\title[Extremely strong C~{\sc iv} $\lambda$1550 in J2229$+$2725]
{Extremely strong C~{\sc iv} $\lambda$1550 nebular emission in the extremely
low-metallicity star-forming galaxy J2229$+$2725}
\author[Y. I. Izotov et al.]{Y. I.\ Izotov$^{1}$\thanks{Corresponding author: yizotov@bitp.kiev.ua},
D. Schaerer$^{2,3}$, 
N. G.\ Guseva$^{1}$,
T. X.\ Thuan$^{4}$
and G. Worseck$^{5}$ \\
                $^{1}$Bogolyubov Institute for Theoretical Physics,
                     National Academy of Sciences of Ukraine,
                     14-b Metrolohichna str., Kyiv, 03143, Ukraine,\\
$^{2}$Observatoire de Gen\`eve, Universit\'e de Gen\`eve, 
51 Ch. des Maillettes, 1290, Versoix, Switzerland,\\
$^{3}$IRAP/CNRS, 14, Av. E. Belin, 31400 Toulouse, France,\\
$^{4}$Astronomy Department, University of Virginia, 
P.O. Box 400325, Charlottesville, VA 22904-4325,\\
$^{5}$ Institut f\"ur Physik und Astronomie, Universit\"at Potsdam, Karl-Liebknecht-Str. 24/25, D-14476 Potsdam, Germany
}
\begin{document}


\pagerange{\pageref{firstpage}--\pageref{lastpage}} \pubyear{2012}

\maketitle

\label{firstpage}

\begin{abstract}
Using {\sl Hubble Space Telescope} ({\sl HST})/Cosmic Origins Spectrograph
(COS) observations of one of the most metal-poor dwarf star-forming galaxies
(SFG) in the local Universe, J2229$+$2725, we have discovered an extremely strong
nebular C~{\sc iv} $\lambda$1549, 1551 emission-line doublet, with an equivalent
width of 43\AA, several times higher than the value observed so far in 
low-redshift SFGs.
Together with other extreme characteristics obtained from optical
spectroscopy (oxygen abundance 12~+~log(O/H) = 7.085$\pm$0.031,
ratio O$_{32}$ = 
$I$([O~{\sc iii}]$\lambda$5007)/$I$([O~{\sc ii}]$\lambda$3727) $\sim$ 53, 
and equivalent width of the H$\beta$ emission line EW(H$\beta$) = 577\AA),
this galaxy greatly increases the range of physical properties for dwarf
SFGs at low redshift and is a likely analogue of the high-redshift
dwarf SFGs responsible for the reionization of the Universe.
We find the ionizing radiation in J2229$+$2725 to be stellar in origin and the
high EW(C~{\sc iv}~$\lambda$1549,1551) to be due to both extreme ionization
conditions and a high carbon abundance, with a corresponding log C/O~=~$-$0.38,
that is $\sim$ 0.4 dex higher than the average value for nearby low-metallicity
SFGs.
\end{abstract}


\section{Introduction}\label{sec:INT}

In a scenario of hierarchical structure formation and during the epoch of the reionization
of the Universe, the majority of galaxies are expected to be of low-mass,
strongly star-forming, and highly
deficient in metals \citep[e.g. ][]{Ma23}.
The low metallicity stellar populations in these galaxies with spectra steeply
rising toward shorter wavelenghts imply strong and possibly hard ionizing
radiation, resulting in some cases in the presence of nebular emission of
high-ionization species such as the C~{\sc iv} $\lambda$1549, 1551 
doublet (hereafter C~{\sc iv} $\lambda$1550) in the UV range
\citep[e.g. ][]{Se17,Be19}.
Furthermore, these galaxies likely dominate the ionizing
background at $z > 6$, and are thus probably the main sources of cosmic
reionization, according to both recent observational evidence from their local
analogues
\citep{I16a,I16b,I18b,I18c,I21a,I22,Fl22} and state-of-the-art numerical
simulations \citep[e.g. ][]{Ro18,Ka21,Ka22}. However,
the detection of these high-redshift galaxies is a difficult task because of
their low mass and hence their extreme faintness. Only recently \citet{A23}, using
{\sl James Webb Space Telescope} ({\sl JWST}) spectroscopic observations, have found
$z$ $\sim$ 6 - 7 galaxies with stellar masses and metallicities as low as those
in the most metal-deficient galaxies at low redshift. However detailed studies of these far-away objects remain difficult.

On the other hand, extensive searches for extremely metal-deficient 
galaxies (XMDs) at low-redshift have revealed star-forming galaxies (SFR) with masses
down to $M_\star$ $\sim$ 10$^6$ M$_\odot$ and metallicities down to
12~+~log(O/H)~$\sim$~7.0 \citep[e.g. ][]{H16,Hsyu17,Hsyu18,SS19,I18a,Iz19a,Iz21,Ko20}.
With these properties and showing very high star-formation activity
(high specific
star-formation rates, sSFR $\sim$ 100 - 300 Gyr$^{-1}$), these XMDs are likely the closest local
analogues of dwarf galaxies in the early Universe. They
represent unique laboratories to study the building blocks of galaxies in so-far
unreachable detail.

\begin{figure}
\centering
\includegraphics[angle=-90,width=0.90\linewidth]{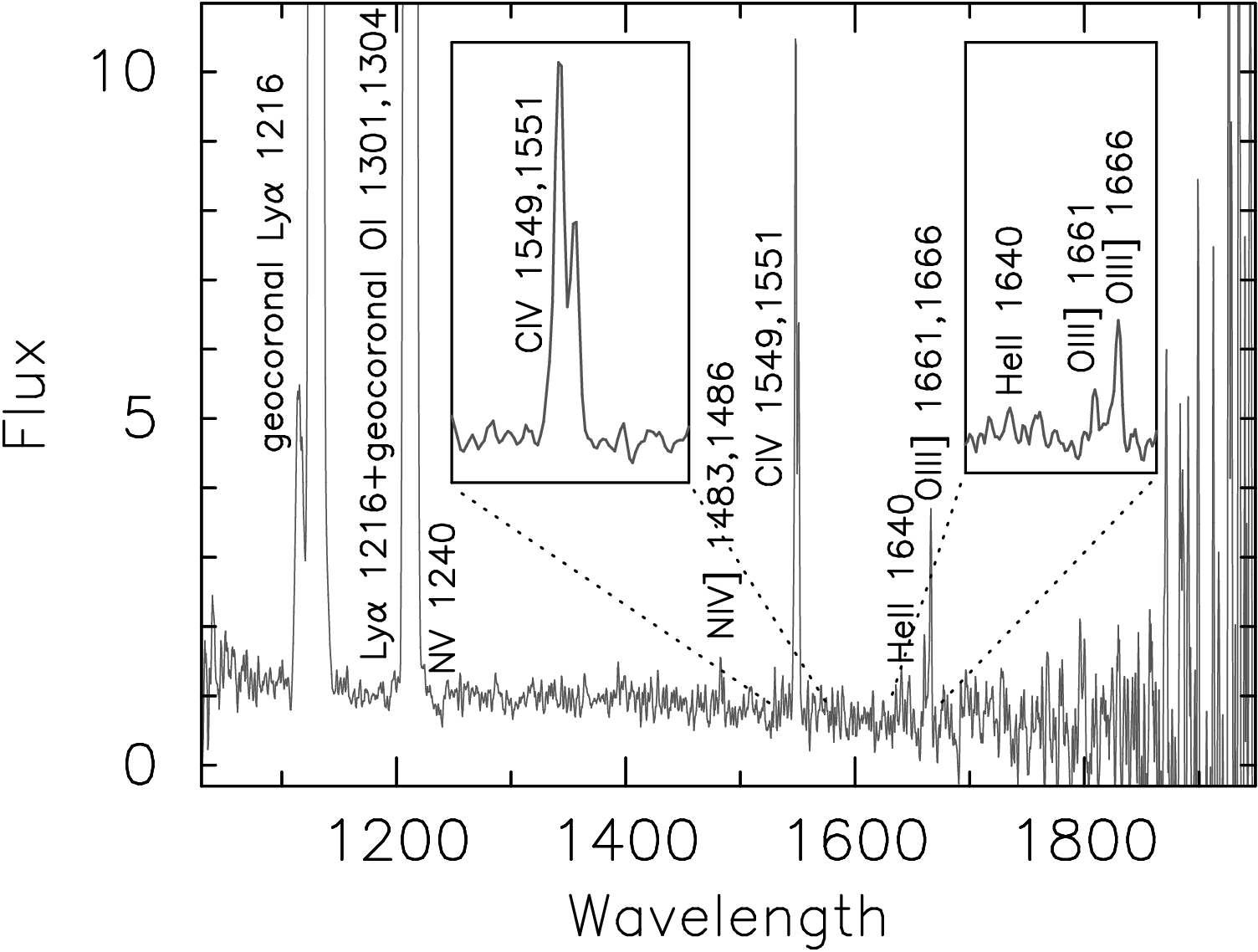}
\caption{The restframe {\sl HST}/COS spectrum of J2229$+$2727 with labelled
lines.
Fluxes are
in units 10$^{-16}$ erg s$^{-1}$ cm$^{-2}$ \AA$^{-1}$ and wavelengths are in \AA.
}
\label{fig1}
\end{figure}

Until recently, most of low-$z$ metal-deficient galaxies were studied in the
rest-frame optical range and very few UV spectra of them have been obtained.
Below 12~+~log(O/H)~$<$~7.5, only few low-redshift star-forming galaxies (SFG) have
been observed with the {\sl Hubble Space Telescope} ({\sl HST}) so far
\citep{Be16,Be18,Be19,Se22}. Recent searches have significantly expanded the
samples of XMDs down to 12 + log(O/H) $\sim$ 7.0, which now
allow us to probe the least chemically-evolved SFGs and study
their UV spectra.
This permits to observe for the first time the main UV emission lines of
such extreme galaxies, in the wavelength range between
Ly$\alpha$ (1216\AA) and $\sim$~1900~\AA, and which include
the C~{\sc iv}~$\lambda$1550 doublet, He~{\sc ii}~$\lambda$1640,
O~{\sc iii}]~$\lambda$$\lambda$1661,~1666
(hereafter O~{\sc iii}]~$\lambda$1663), and possibly other lines.

In this paper we present {\sl HST} observations of one of the most
metal-deficient galaxies, J2229$+$2725, with 12~+~log(O/H)~=~7.085. Its properties
from optical spectroscopic observations were studied in detail by
\citet{Iz21}. Some of these properties are shown in Table~\ref{tab1}.
The most notable are an extremely high equivalent width EW(H$\beta$) of 577\AA\
and an extremely high O$_{32}$ ratio of 53. Both characteristics indicate that
the ionizing parameter and/or the hardness of ionizing radiation are considerably
higher in J2229$+$2725 than in other star-forming galaxies.

\section{The UV spectrum of J2229$+$2725}\label{sec:observations}

{\sl HST}/Cosmic Origins Spectrograph (COS) spectroscopy of J2229$+$2725 was
obtained in program GO 17100 (PI:~Y.\,I.\,Izotov) on 23 July 2023. The galaxy
was acquired by COS near-ultraviolet (NUV) imaging.
The spectrum with an exposure time of 12282 sec was obtained with the
G140L grating. It was reduced with the CALCOS pipeline v3.3.4.

The UV spectrum of J2229$+$2725 is shown in Fig.~\ref{fig1}. Its most striking
feature is the very strong narrow
C~{\sc iv}~$\lambda$1550 doublet, without any indication of the
stellar absorption lines which are commonly seen in star-forming galaxies
with higher metallicities. The weaker nebular lines
N~{\sc iv}]~$\lambda$1483,~1486, He~{\sc ii}~$\lambda$1640, and
O~{\sc iii}]~$\lambda$1663 doublet are also detected. We also note the presence
of the stellar absorption line N~{\sc v}~$\lambda$1240, produced by massive
stars with stellar winds. Unfortunately, the C~{\sc iii}]~$\lambda$1909
emission line falls into the noisy part of the spectrum and therefore could not
be seen.

\begin{table}
\centering
\caption{Observed and derived characteristics of J2229$+$2725$^\dag$ \label{tab1}}
\begin{tabular}{lr} \hline
Parameter                 &  J2229$+$2725       \\ \hline
R.A.(J2000)               &  22:29:33.19 \\
Dec.(J2000)               & +27:25:25.60 \\
  $z$                     &  0.07622     \\
 $M_g$, mag         & $-$16.39$\pm$0.06     \\
log $L_g$/L$_{g,\odot}$&     8.74     \\
log $M_\star$/M$_\odot$&   6.96       \\
$L$(H$\beta$), erg s$^{-1}$&(3.1$\pm$0.3)$\times$10$^{40}$\\
EW(H$\beta$), \AA\       &   577$\pm$4 \\    
SFR, M$_\odot$yr$^{-1}$  &     0.68$\pm$0.06 \\
sSFR, Gyr$^{-1}$          &     75 \\
O$_{32}$                  &     53 \\
$T_{\rm e}$(O {\sc iii}), K          &   24800$\pm$900       \\
$N_{\rm e}$(S {\sc ii}), cm$^{-3}$    &1000$\pm$600         \\
  12+logO/H              &7.085$\pm$0.031 \\
\hline
  \end{tabular}


\noindent$^\dag$Data are taken from \citet*{Iz21}.

  \end{table}

\begin{figure}
\centering
\includegraphics[angle=-90,width=0.90\linewidth]{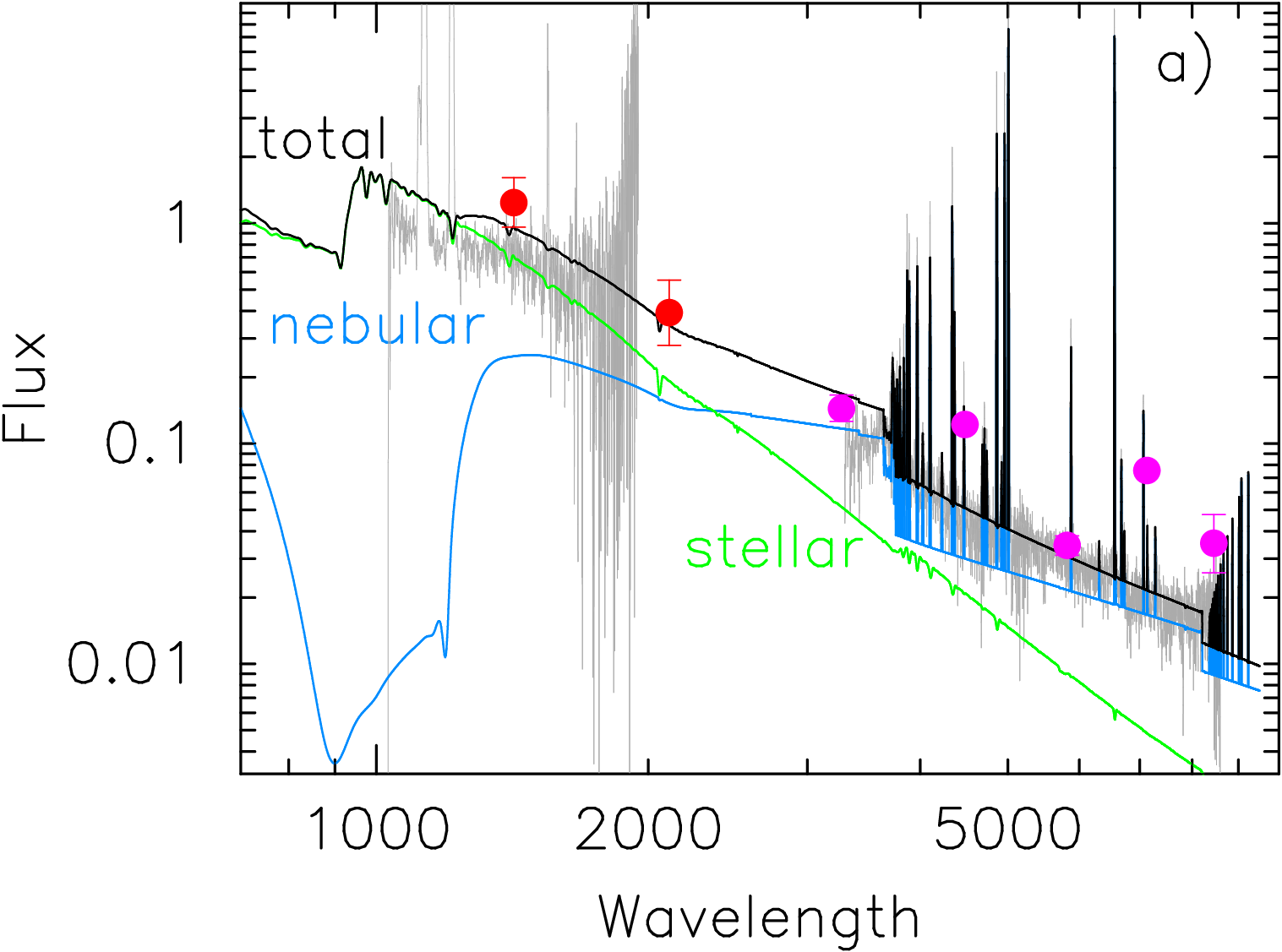}
\vspace{0.2cm}
\caption{
The SED obtained from ﬁtting the rest-frame
Large Binocular Telescope (LBT) spectrum taken from \citet{Iz21}. The observed
UV and optical spectra are shown in grey, the modelled stellar, nebular, and
total SEDs are shown by green, blue, and black lines, respectively, and are
labelled. {\sl GALEX} FUV, NUV, and SDSS $u, g, r, i$, and $z$ photometric data,
blueshifted adopting the J2229$+$2725 redshift, are shown by red and magenta
ﬁlled circles, respectively. Fluxes are
in units 10$^{-16}$ erg s$^{-1}$ cm$^{-2}$ \AA$^{-1}$ and wavelengths are in \AA.
}
\label{fig2}
\end{figure}

In Fig.~\ref{fig2} we show the COS UV and Large Binocular Telescope (LBT)
optical spectra of J2229$+$2725, superposed
on the spectral energy distribution (SED) (black line) derived by \citet{Iz21}
from the LBT spectrum and extrapolated to the UV range. The SED consists of
two components, nebular and stellar (blue and green lines, respectively).
It is seen \citep[as already noted by ][]{Iz21} that the contribution of
the nebular emission to the SED is high even in the UV range and dominates
in the optical range. This is because of the extremely high EW(H$\beta$).
The SED of the nebular continuum includes two-photon hydrogen emission and
free-free and free-bound emission of hydrogen and helium \citep{A84}.

\begin{table}
\caption{Extinction-corrected line fluxes and equivalent widths in the UV spectrum of J2229$+$2725 \label{tab2}}
\begin{tabular}{lrrr} \hline
Line                 &\multicolumn{1}{c}{$F$$^*$}&  \multicolumn{1}{c}{100$\times$$I$($\lambda$)/$I$(H$\beta$)$^\dag$}& \multicolumn{1}{c}{EW$^\ddag$}     \\ \hline
N~{\sc v} $\lambda$1240    &$-$2.30$\pm$0.40&$-$13.92$\pm$2.91  & $-$3.2$\pm$0.7 \\
N~{\sc iv}] $\lambda$1483  & 0.61$\pm$0.58&   3.27$\pm$3.12 &    1.1$\pm$1.0 \\
N~{\sc iv}] $\lambda$1486  & 0.44$\pm$0.58&   2.38$\pm$3.12 &    0.9$\pm$1.1 \\
C~{\sc iv} $\lambda$1549   &12.43$\pm$1.22&  66.81$\pm$6.53 &   26.6$\pm$2.7 \\
C~{\sc iv} $\lambda$1551   & 7.00$\pm$1.21&  37.63$\pm$6.50 &   16.2$\pm$2.8 \\
He~{\sc ii} $\lambda$1640  & 0.97$\pm$0.45&   5.15$\pm$2.41 &    2.6$\pm$1.2 \\
O~{\sc iii}] $\lambda$1661 & 1.29$\pm$0.58&   6.87$\pm$3.12 &    3.3$\pm$1.2 \\
O~{\sc iii}] $\lambda$1666 & 3.28$\pm$0.58&  17.46$\pm$3.12 &    9.1$\pm$1.2 \\
\hline
  \end{tabular}


\noindent$^*$$F$ is the observed flux in units of 10$^{-16}$ erg s$^{-1}$cm$^{-2}$.

\noindent$^\dag$$I$(H$\beta$) = 24.63$\times$10$^{-16}$ erg s$^{-1}$ cm$^{-2}$
\citep{Iz21} and $I$($\lambda$) are extinction-corrected fluxes.

\noindent$^\ddag$Equivalent width in \AA.

  \end{table}

Nebular emission in the continuum of this galaxy is considerable even in the UV 
range longward of the Ly$\alpha$ line, due to sharply rising two-photon emission
at $\lambda$~$\geq$~1215.67~\AA. Its contribution should be taken into account
in the determination of stellar masses and of the UV stellar slope $\beta$,
commonly used in studies of 
star-forming galaxies. Neglecting the nebular continuum contribution would
make the stellar mass larger and the UV slope shallower.

We also note that the observed UV spectrum in Fig.~\ref{fig2} is in fair
agreement with the extrapolated SED. However, there is still some offset
present. This offset can be caused by various reasons, for example, by
uncertainties of SED fitting or possible vignetting of the COS spectrum.

We measure the line fluxes and
equivalent widths in the observed UV spectrum and correct them for extinction,
adopting $C$(H$\beta$) = 0.060$\pm$0.038 \citep{Iz21} and the reddening law
of  \citet*{C89} with $R(V)$ = 2.7. The results of the measurements are shown in
Table~\ref{tab2}.

\section{Origin of strong nebular C~{\sc iv} $\lambda$1550 emission}

It is seen from Table~\ref{tab2} that the equivalent width of the
C~{\sc iv}~$\lambda$1550 nebular emission doublet is extremely high,
$\sim$~43\AA. Such a high value is detected for the first time in the
low-redshift and low-mass SFGs. However, it is similar to EW(C~{\sc iv}) $\sim$
38\AA\ found in the high-$z$ galaxy A1703-zd6 at $z$ = 7.045 \citep{St15}.
We also derive the high ionizing photon
production efficiency $\xi_{\rm ion}$ =~10$^{25.92}$~Hz~erg$^{-1}$, using 
the extinction-corrected H$\beta$ emission-line flux and the
extinction-corrected flux of the COS spectrum at the rest-frame wavelength
of 1500\AA\ 
(Fig.~\ref{fig2}). These facts together with other extreme characteristics
derived by \citet{Iz21} make J2229$+$2725 an important marker, 
indicating the possible extended range of physical characteristics and
tendencies in their relationships for galaxies with very young starbursts.
The extremely high O$_{32}$ = 53 of J2229$+$2725 corresponds to a very high
ionization parameter $U$ $\sim$ 10$^{-1.7}$, according to \citet{Be19}. This high
$U$ is due to an intense
ionizing radiation, a young starburst age, corresponding to EW(H$\beta$) =
577\AA, and a very compact structure of J2229$+$2725 \citep{Iz24}, with a high
electron number density in its H~{\sc ii} region
$N_{\rm e}$([S~{\sc ii}]) $\sim$ 10$^3$ cm$^{-3}$ and
$N_{\rm e}$([Ar~{\sc iv}]) $\sim$ (3-4)$\times$10$^3$ cm$^{-3}$ \citep{Iz21}.

However, it is likely  
that the H~{\sc ii} region is density-bounded. This leads to a reduction of the
O$^+$ zone and to an increase of O$_{32}$. Indeed, the
double-peaked Ly$\alpha$ emission line is narrow in J2229$+$2725, with a small
peak velocity separation of 207 km s$^{-1}$, corresponding to an escape fraction
of ionizing LyC emission $f_{\rm esc}$(LyC) $\sim$ 34 per cent \citep{Iz24}.

\begin{figure*}
\hbox{
\includegraphics[angle=-90,width=0.32\linewidth]{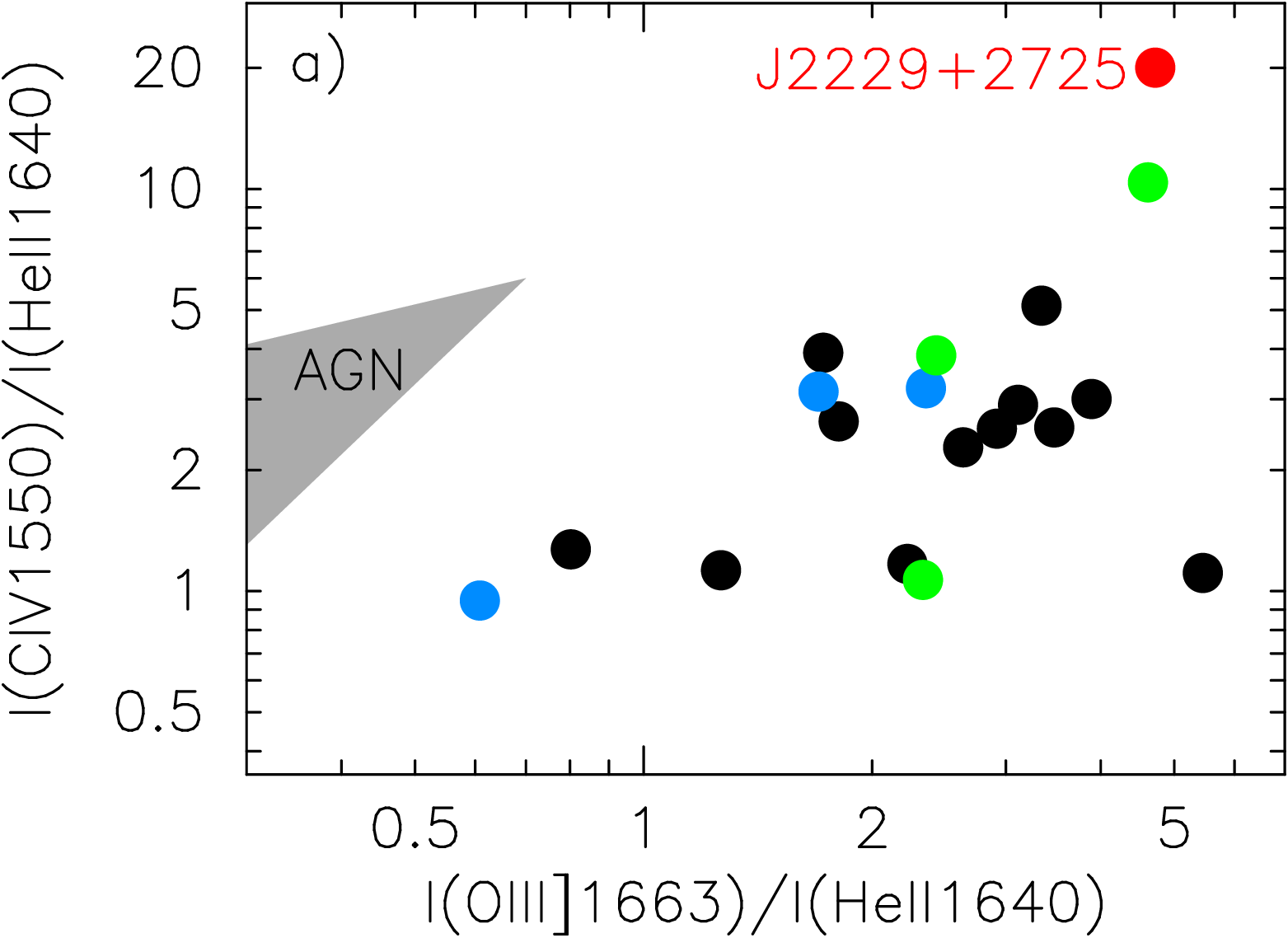}
\hspace{0.1cm}\includegraphics[angle=-90,width=0.32\linewidth]{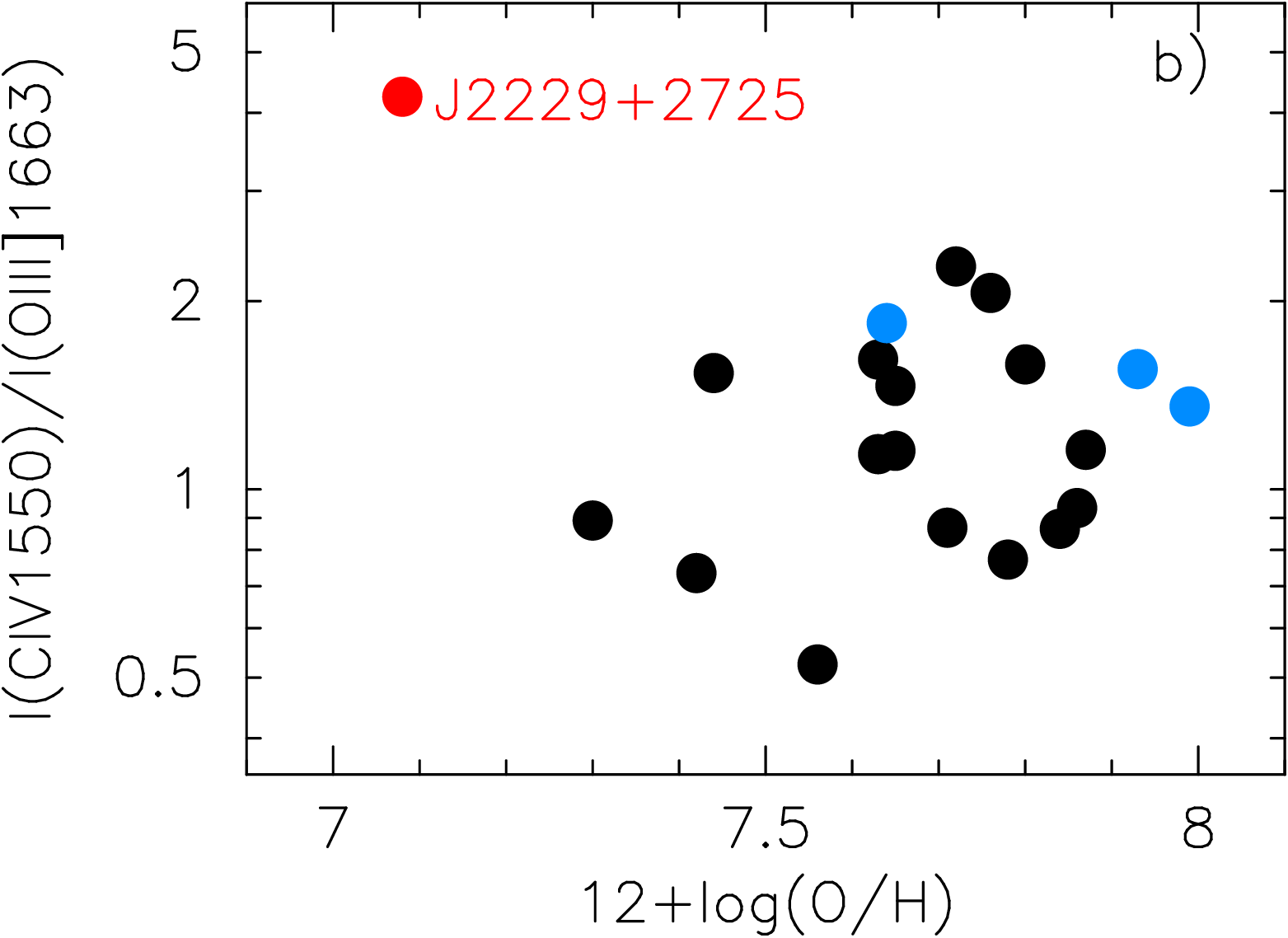}
\hspace{0.1cm}\includegraphics[angle=-90,width=0.32\linewidth]{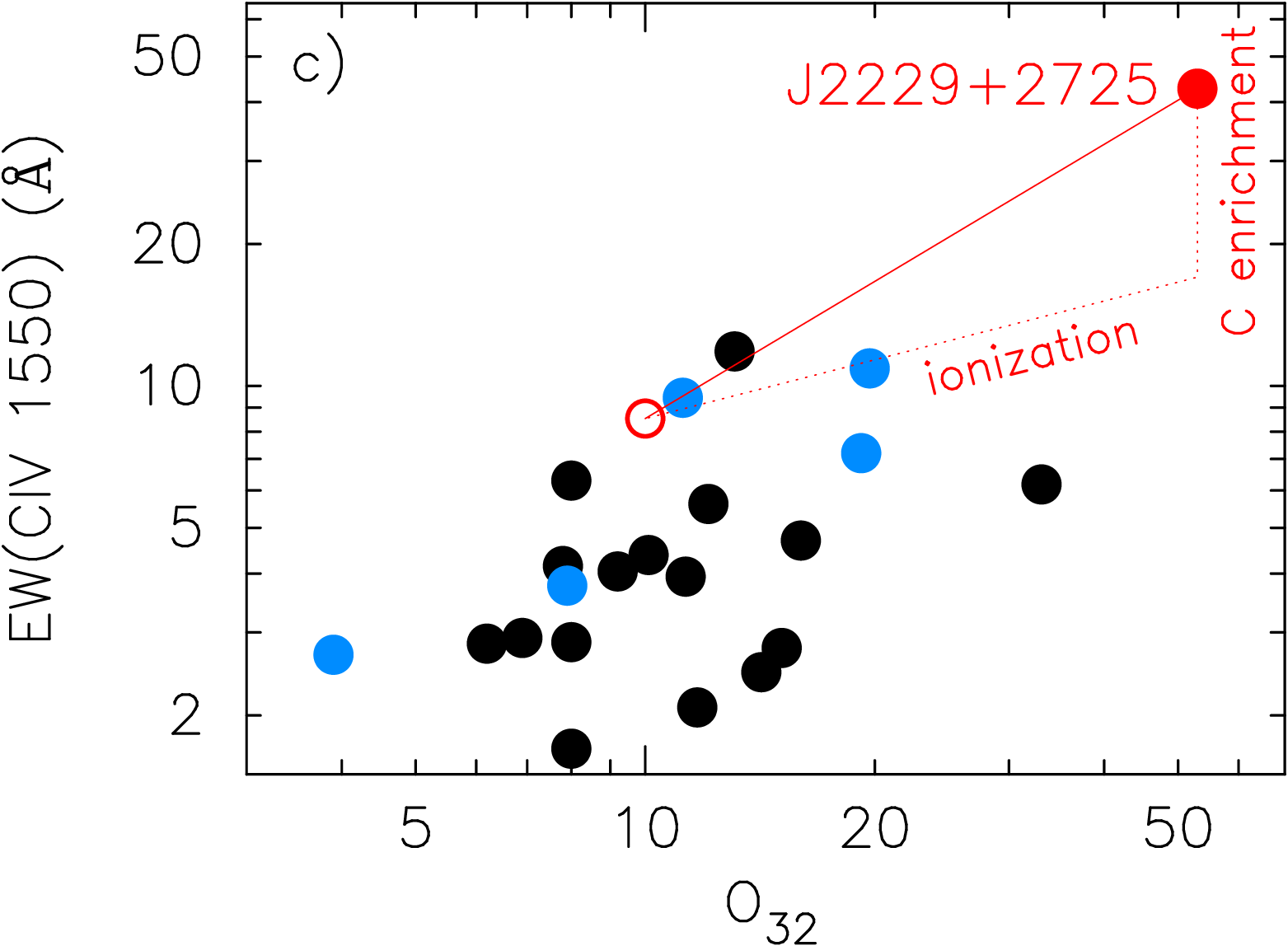}
}
\vspace{0.2cm}
\hbox{
\includegraphics[angle=-90,width=0.32\linewidth]{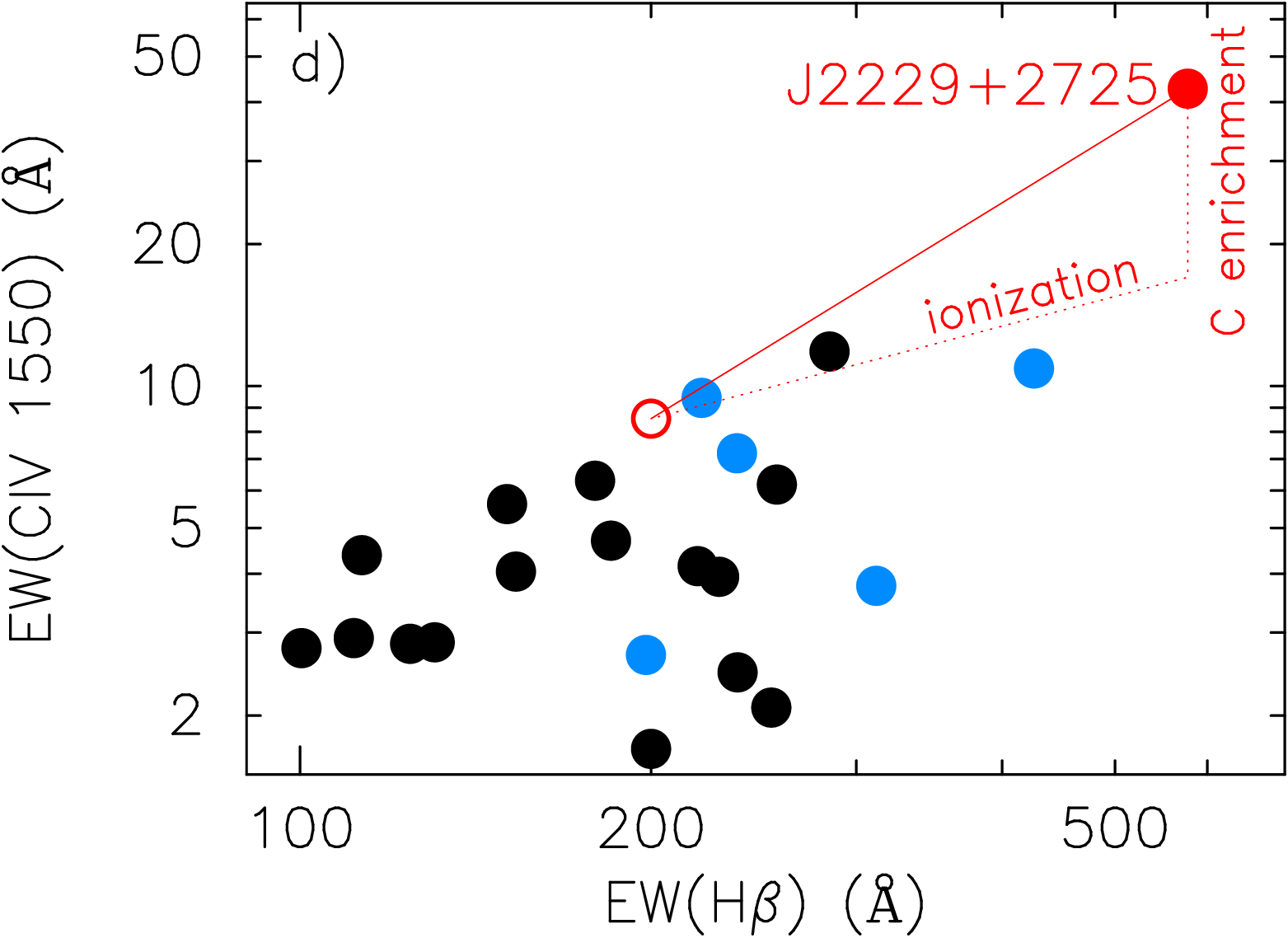}
\hspace{0.1cm}\includegraphics[angle=-90,width=0.32\linewidth]{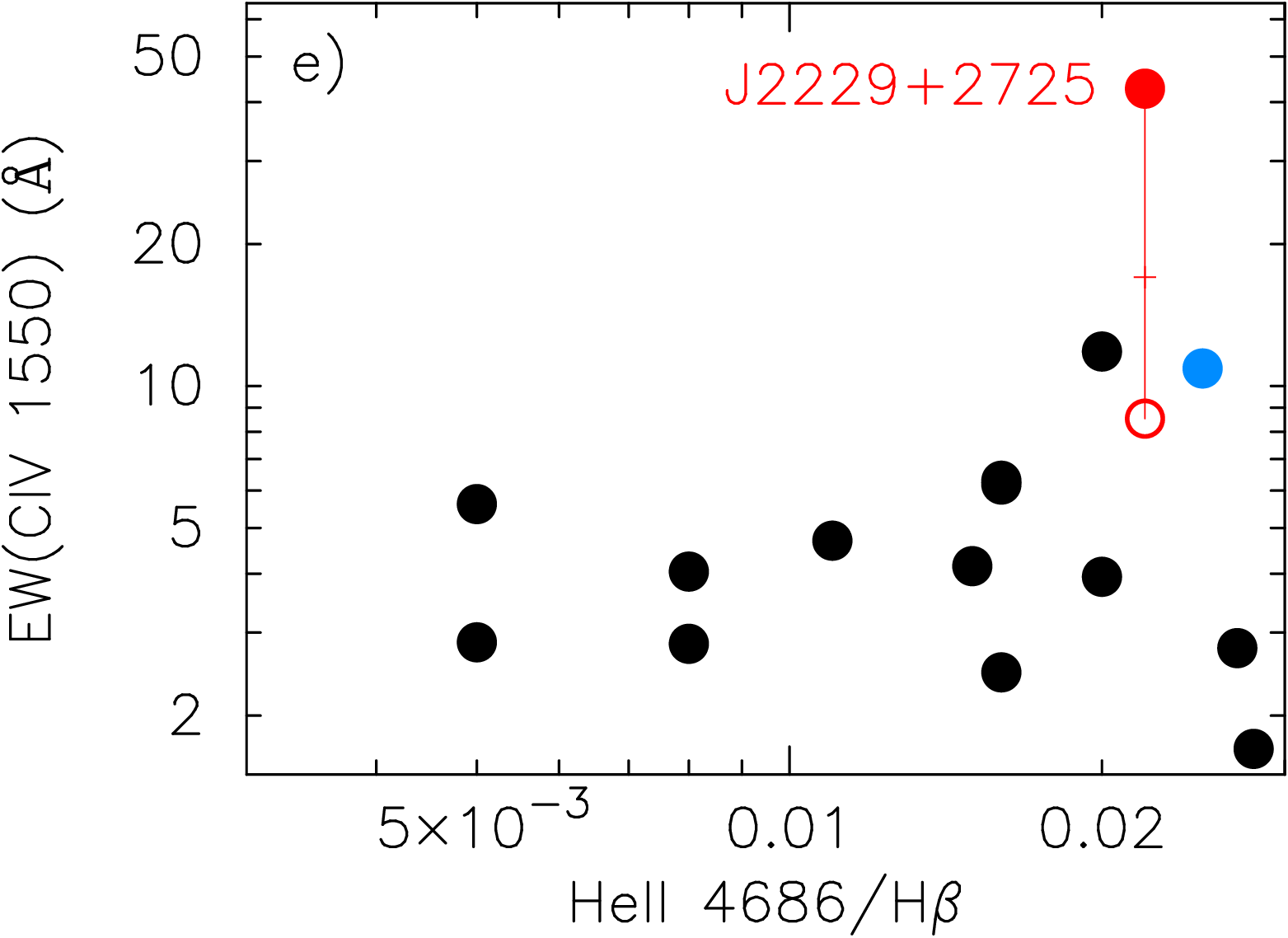}
\hspace{0.1cm}\includegraphics[angle=-90,width=0.32\linewidth]{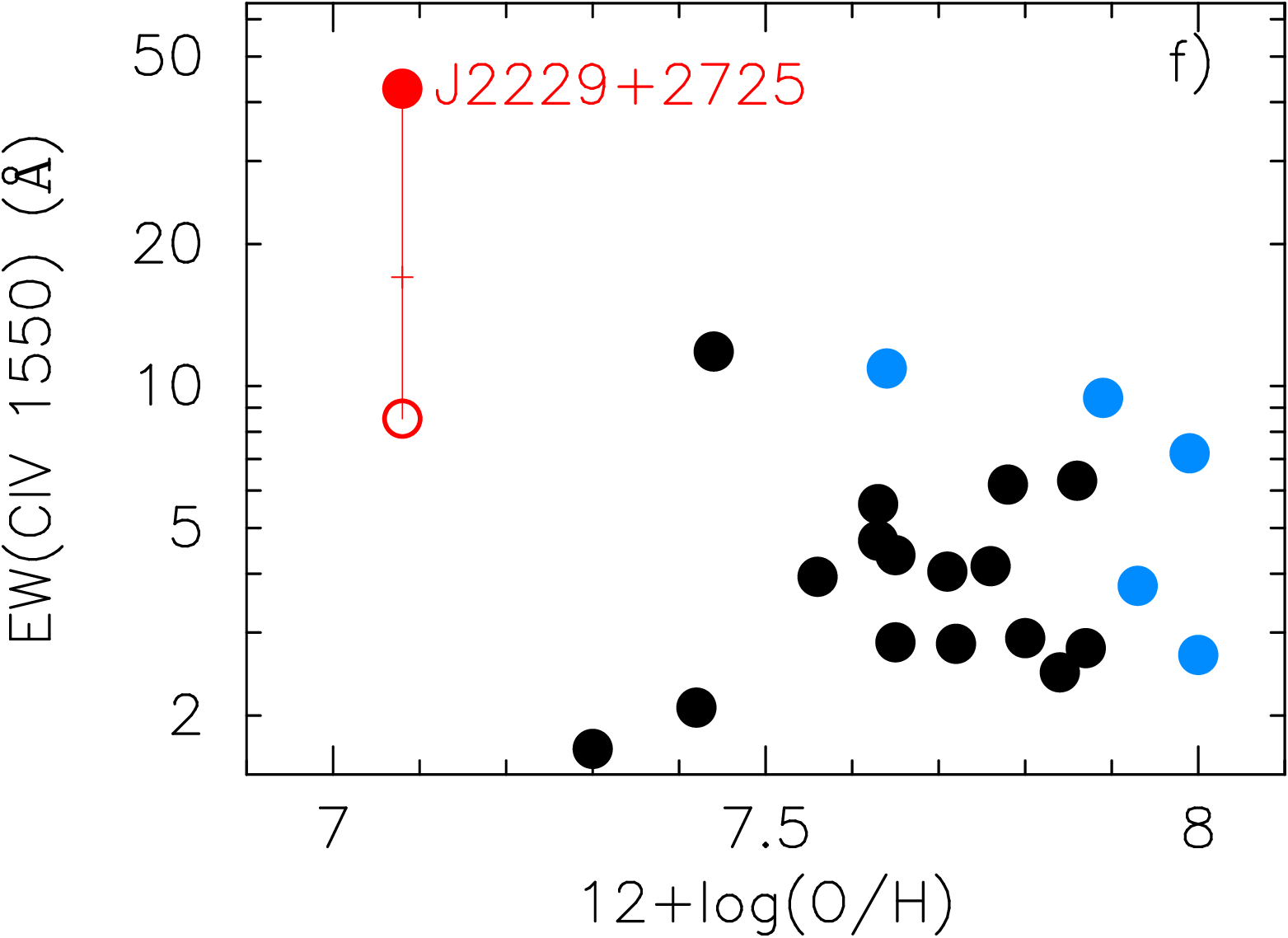}
}
\caption{(a) Diagnostic diagram
$I$(C~{\sc iv}\,$\lambda$1550)/$I$(He~{\sc ii}\,$\lambda$1640)
-- $I$(O~{\sc iii}]\,$\lambda$1663)/$I$(He~{\sc ii}\,$\lambda$1640). The
location of AGN \citep{Ta21} is shown by the shaded grey region and labelled.
(b) The relation between C~{\sc iv}\,$\lambda$1550/O~{\sc iii}]\,$\lambda$1663
flux ratio and oxygen abundance. (c) - (f) Relations
between the equivalent widths EW(C~{\sc iv}$\,\lambda$1550) and (c) the
O$_{32}$ ratios, (d) the equivalent widths EW(H$\beta$) of the H$\beta$ emission
line, (e) the He~{\sc ii}\,$\lambda$4686/H$\beta$ flux ratio,
and (f) the oxygen abundances 12+log(O/H).
In all panels, J2229+2725 is shown by a red filled circle,
low-redshift ($z$ $<$ 0.1) galaxies from \citet{Be16,Be19}, \citet{Se19},
\citet{W21} by black circles and
$z$ $\sim$ 0.3 LyC leaking galaxies \citep{Sc22} by blue circles. High-redshift
galaxies by \citet{V16,V17,V20} in (a) are represented by green filled circles.
Red open circles in (c) -- (f) show the tentative location of J2229$+$2725
if it would have typical characteristics of the galaxies from the comparison sample,
i.e. log(C/O)~=~$-$0.75, O$_{32}$ $\sim$ 10 and EW(H$\beta$) $\sim$ 200\AA.
\label{fig3}
}
\end{figure*}

We show in Fig.~\ref{fig3}a the diagnostic diagram
$I$(C~{\sc iv} $\lambda$1550)/$I$(He~{\sc ii} $\lambda$1640) --
$I$(O~{\sc iii}] $\lambda$1663)/$I$(He~{\sc ii} $\lambda$1640)
to constrain the origin of the ionizing radiation responsible for the strong
C~{\sc iv}~$\lambda$1550 emission. J2229$+$2725 is located at
the upper end of the distribution of low- and high-$z$ SFGs,
far from the location of AGN (shaded region), indicating that ionizing radiation
in this galaxy producing C~{\sc iv} $\lambda$1550 is stellar in origin.
The absence of an AGN is to be expected for this galaxy 
as it is fully compatible with the optical line ratios,
which clearly indicate stellar photoionization \citep[see ][]{Iz21}.

In Fig.~\ref{fig3}b, we show the relation between
the C~{\sc iv}~$\lambda$1550/O~{\sc iii}]~$\lambda$1663 flux ratio and the oxygen
abundance 12+log(O/H) for SFGs.
This ratio in J2229$+$2725 (filled red circle) is considerably higher than
those for SFGs from a comparison sample (blue and black filled circles).
We consider two
possible causes for this appearance: 1) a higher C/O abundance ratio
compared to those in SFGs from the comparison sample and 2) higher
ionization efficiency in this compact and dense galaxy, indicated by its
extremely high O$_{32}$ ratio, which is an observable characteristic of the high
ionization parameter $U$, and very high $\xi_{\rm ion}$. Both these effects may result in an increase of
the C~{\sc iv}~$\lambda$1550/O~{\sc iii}]~$\lambda$1663 flux ratio and of the
equivalent width of C~{\sc iv}~$\lambda$1550 doublet in J2229$+$2725.

\begin{figure}
\centering
\includegraphics[angle=-90,width=0.90\linewidth]{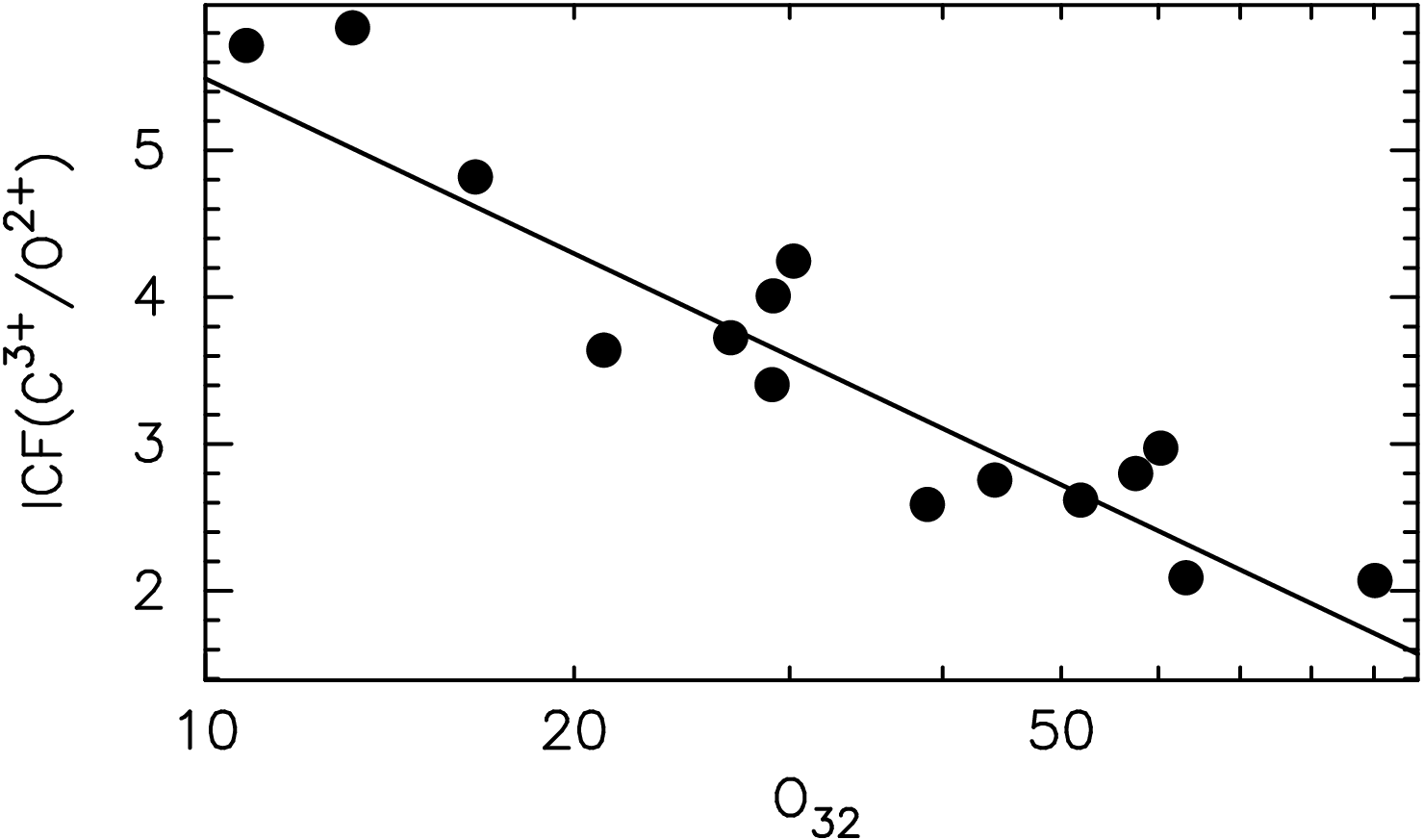}
\caption{Relation between the ionization correction factor
ICF(C$^{3+}$/O$^{2+}$) and the O$_{32}$ ratio for O$_{32}$ $\geq$ 10
shown by circles and obtained by \citet{Iz23} using the Cloudy code of
\citet{F13}. The maximum likelihood regression
to the data (Eq.~\ref{ICF}) is shown by a solid line.}
\label{fig4}
\end{figure}

We adopt equation 4 of \citet{PMA17} and the electron temperature
$T_{\rm e}$(O~{\sc iii})~=~24800~K (Table~\ref{tab1}) for both the C$^{3+}$ and O$^{2+}$ species to derive the
C$^{3+}$/O$^{2+}$ abundance ratio, using the extinction-corrected flux ratios of
C~{\sc iv}~$\lambda$1550 and O~{\sc iii}]~$\lambda$1663 (Table~\ref{tab2}).
We obtain log~(C$^{3+}$/O$^{2+}$)~=~$-$0.76.

The total C/O abundance is derived from Eq.~\ref{co}, using the ionization
correction factor ICF(C$^{3+}$/O$^{2+}$):
\begin{equation}
\frac{\rm C}{\rm O} = \frac{{\rm C}^{3+}}{{\rm O}^{2+}} \times {\rm ICF} \left( \frac{{\rm C}^{3+}}{{\rm O}^{2+}} \right). \label{co}
\end{equation}
For the determination of ICF(C$^{3+}$/O$^{2+}$),
we consider a set of the same photoionized H~{\sc ii} region models 
calculated with the code {\sc cloudy} \citep{F13}
as the ones used by \citet{Iz23}. However, here we select only models with an
young age of 1 Myr, a high number density $N_{\rm e}$ = 10$^3$ cm$^{-3}$ and
O$_{32}$ $>$ 10. These parameters closely represent the physical conditions in
J2229$+$2725 \citep{Iz23}. The relation between ICF(C$^{3+}$/O$^{2+}$) and
O$_{32}$ in these models is shown in Fig.~\ref{fig4} and can be fit by the
maximum likelihood linear regression
\begin{equation}
{\rm ICF} \left( \frac{{\rm C}^{3+}}{{\rm O}^{2+}} \right) = 9.45-3.96\times \log {\rm O}_{32}. \label{ICF}
\end{equation}
We obtain ICF(C$^{3+}$/O$^{2+}$) = 2.62 (or log ICF(C$^{3+}$/O$^{2+}$) = 0.42 dex),
adopting the O$_{32}$ value of 53 observed in
J2229$+$2725. Finally, from Eq.~\ref{co} we derive log(C/O)~=~$-$0.38.
This log(C/O) value is $\sim$ 0.4 dex higher than the average value log(C/O)
of $\sim$ $-$0.75 for low-metallicity SFGs \citep[e.g. ][]{Iz23}.
Using Cloudy models, we can estimate the fractions of C$^{+2}$ and C$^{+3}$
in the H~{\sc ii} region of J2229$+$2725. We find that the fraction of
C$^{+2}$ in models, predicting O$_{32}$ $\sim$ 50 -- 60, is higher by a factor of
$\sim$ 1.5. Then log(C/O) $\sim$
log[(C$^{+2}$+C$^{+3}$)/O$^{+2}$] = log(C$^{+3}$/O$^{+2}$) + 0.4 = --0.76 + 0.4
= --0.36, or roughly similar to --0.38 derived from log(C$^{+3}$/O$^{+2}$).

However, the O$_{32}$ ratio can be smaller, because of the decrease of the O$^+$
zone by a factor of $\sim$ 1.35. This results in O$_{32}$ $\sim$ 39 if the LyC
leakage is taken into account. But in this case, log (C/O) would be further
increased by $\sim$ 0.06 dex because of an increasing ICF (Fig.~\ref{fig4}).

Additionally, the equivalent width of the C~{\sc iv} $\lambda$1550 doublet
can also increase due to extreme ionization conditions in J2229$+$2725.
This is demonstrated by the relations
EW(C~{\sc iv} $\lambda$1550) -- O$_{32}$
(Fig.~\ref{fig3}c) and EW(C~{\sc iv} $\lambda$1550) -- EW(H$\beta$)
(Fig.~\ref{fig3}d). In both panels, EW(C~{\sc iv}~$\lambda$1550) increases with
increasing O$_{32}$ and EW(H$\beta$), respectively, with J2229$+$2725 having the
highest values. Its extremely high O$_{32}$ = 53 corresponds to very intense
ionizing radiation compared to other SFGs.
A similar conclusion can be drawn from the equivalent width of the H$\beta$
emission line which is a characteristic of the starburst age. Its
extremely high EW(H$\beta$) = 577\AA\ indicates a very young age of
$\sim$ 1 -- 3 Myr \citep{L99} for the 
starburst in J2229$+$2725. Thus, the top location of J2229$+$2725 in both
relations leads us to conclude that it is the more intense ionizing radiation in
this galaxy that is responsible for its higher EW(C~{\sc iv}~$\lambda$1550).

Where would J2229$+$2725 be located in Fig.~\ref{fig3}c and
\ref{fig3}d if its characteristics are similar to those in galaxies
from the comparison sample, i.e. log(C/O)~=~$-$0.75, O$_{32}$ $\sim$ 10 and
EW(H$\beta$) $\sim$ 200\AA? In this case, EW(C~{\sc iv} $\lambda$1550) should
decrease by a factor of $\sim$~2.5 due to the correction for the carbon
overabundance. Furthermore, the ionization correction factor
ICF(C$^{3+}$/O$^{2+}$) at O$_{32}$ $\sim$ 10, with less intense ionizing
radiation, is a factor of $\sim$~2 higher than the one at the observed value
O$_{32}$~=~53 (Fig.~\ref{fig4}). This means that the fraction of carbon in the
C$^{3+}$ form at a fixed C/O abundance ratio would be lower by the same factor
whereas the fraction of oxygen in the O$^{2+}$ form at O$_{32}$ $\geq$ 10
remains nearly constant.
Thus, EW(C~{\sc iv} $\lambda$1550) would be reduced by a factor of $\sim$~5
due to both effects, attaining a value of $\sim$~8\AA, typical of the
SFGs from the comparison sample. The reduced value is shown by a red open circle
in Fig~\ref{fig3} c-f.

The ionizing radiation may also be harder at low metallicities.
However, no correlation is found between the
He~{\sc ii}~$\lambda$4686/H$\beta$ flux ratio and EW(C~{\sc iv} $\lambda$1550)
(Fig.~\ref{fig3}e) for galaxies from the comparison sample.
The galaxy J2229$+$2725 is off this distribution, but correcting
EW(C~{\sc iv}~$\lambda$1550) for enhanced carbon abundance and ionization
conditions would move it to the region occupied by other galaxies.
Furthermore, the He~{\sc ii}~$\lambda$4686/H$\beta$ flux ratio in this galaxy
is in the range of the values observed in galaxies from the comparison sample.
Similarly, no correlation between 12+log(O/H) and
EW(C~{\sc iv} $\lambda$1550) is found for galaxies from the comparison sample
(blue and black symbols in Fig.~\ref{fig3}f). Again, the galaxy J2229$+$2725 is off this
distribution, being the most-metal poor galaxy with an unusually high
EW(C~{\sc iv}~$\lambda$1550). After its correction for the two mentioned above
effects, we find no correlation, indicating that the ionizing radiation in
J2229$+$2725 with very low metallicity is likely not harder than in other
galaxies.

If confirmed by other carbon abundance determinations, e.g. using
the C~{\sc iii}] $\lambda$1909 emission line, the origin of the
enhanced C/O abundance ratio in J2229$+$2725 remains to be explained.
On the other hand, the N/O abundance ratio is normal for its low metallicity
\citep{Iz21}.

\section{Conclusions}\label{sec:conclusions}

We present {\sl Hubble Space Telescope} ({\sl HST})/Cosmic Origins Spectrograph
(COS) spectrophotometric observations of the compact star-forming galaxy (SFG)
J2229$+$2725. This galaxy possesses extraordinary properties, as derived from
earlier optical observations. In particular, it is characterized by a very low
oxygen abundance 12~+~log(O/H)~=~7.085, an extremely high O$_{32}$ ratio of 53
and an extremely high equivalent width EW(H$\beta$)~=~577\AA.

We find the properties of this galaxy in the
UV to be no less unusual. We discover an extremely strong nebular
C~{\sc iv}~$\lambda$1550 doublet in emission with an equivalent width of 43\AA.
This is several times higher than in other low-$z$ star-forming galaxies
with detected C~{\sc iv}~$\lambda$1550 nebular emission. The detection
of such a strong line, together with some other extreme properties of
J2229$+$2725, makes this galaxy an excellent marker, allowing to enlarge the
range of characteristics which SFGs at low- and high-redshifts may possess.

We also find that the ionizing radiation in J2229$+$2725 is stellar in origin. We show that 
its high EW(C~{\sc iv} $\lambda$1550) can be explained by a high C/O abundance
ratio, $\sim$~2.5 times higher than the average value for
low-metallicity SFGs, and by extreme ionization conditions, characterized by a high
O$_{32}$ ratio. These ionization conditions 
further push EW(C~{\sc iv} $\lambda$1550) up by a factor of $\sim$2.

Because of its very low metallicity and stellar mass and exceptionally high
star-formation activity, J2229$+$2725 is likely among the best local analogues
of star-forming dwarf galaxies in the early universe.  
 The results presented here will thus help to shed light on studies of high-redshift dwarf
star-forming galaxies thought to be main source of the reionization of
the Universe.

\section*{Acknowledgements}

These results are based on observations made with the NASA/ESA 
{\sl Hubble Space Telescope}, 
obtained from the data archive at the Space Telescope Science Institute. 
STScI is operated by the Association of Universities for Research in Astronomy,
Inc. under National Aeronautics and Space Administration (NASA) contract NAS 5-26555.
 T.X.T. was supported by 
NASA through grant number HST-GO-17100.002-A from the Space Telescope Science 
Institute.
Y.I.I. and N.G.G. acknowledge support from the National Academy of Sciences of 
Ukraine by its project No. 0123U102248
``Properties of the matter at high
energies and in galaxies during the epoch of the reionization of the Universe''
and from the Simons Foundation.

\section*{Data availability}

The data underlying this article will be shared on reasonable request to the 
corresponding author.

\bsp

\label{lastpage}

\end{document}